\documentstyle[aps,preprint,psfig]{revtex}
\tightenlines

\global\firstfigfalse

\begin{document}


\draft


\title{Diagnostic criterion for crystallized beams}
\author{Harel Primack 
          \footnote{Email: harel@phyc1.physik.uni-freiburg.de}
          and Reinhold Bl\"umel
          \footnote{Email: blumel@phyc1.physik.uni-freiburg.de}}
\address{Fakult\"at f\"ur Physik,
         Albert-Ludwigs-Universit\"at, 
         Hermann-Herder-Str.\ 3,
         D-79104 Freiburg, Germany}
\date{\today}
\maketitle

\begin{abstract}
%
Small ion crystals in a Paul trap are stable even in the absence of
laser cooling. Based on this theoretically and experimentally
well--established fact we propose the following diagnostic criterion
for establishing the presence of a crystallized beam: Absence of
heating following the shut--down of all cooling devices. The validity
of the criterion is checked with the help of detailed numerical
simulations.
%
\end{abstract}




%
The production of crystallized beams is the holy grail of storage ring
physics \cite{SK85,RS86}. As a matter of fact crystallization has
already been achieved in a miniature storage ring with essentially
stationary ions \cite{BKW92}. Crystallization in high energy storage
rings, however, remains an elusive goal although many laboratories
throughout the world are working on the problem (see, e.g.,
\cite{HG95,MGGHS96,HKNPS91,HNPSS95}). Apart from electron 
cooling \cite{Dan95}, laser cooling \cite{HS75,Ste86} is now employed
by some groups \cite{HG95,MGGHS96,HKNPS91,HNPSS95} as the most
efficient and most promising method to achieve the low beam
temperatures required for beam crystallization. Judging from the
enormous progress achieved in the past few years success seems
imminent. But while it is straightforward to observe ion crystals in
traps \cite{Gho95} directly with the help of CCD cameras
\cite{HDRB88,DPCQW87,WBIBM87,BCPQS88,BKQW89}, it is very difficult to
observe the internal structure of a fast beam in a storage ring
directly with optical means. As a matter of fact, no such observation
has been reported so far. Other imaging methods may require structural
changes of existing storage rings which are both technically difficult
and costly. In view of these difficulties it becomes imperative to
devise diagnostic tools different from direct optical observation that
are capable of distinguishing between a crystallized and a
non-crystallized beam. It is the purpose of this paper to suggest a
simple diagnostic criterion capable of making this distinction. The
criterion is based on a hysteresis effect first described
experimentally in connection with ion crystals in a Paul trap
\cite{DPCQW87}. It is rooted in the observation that once an ion crystal
has been produced, e.g., by laser cooling, it remains stable even in
the absence of laser cooling \cite{BKQW89}. Since in the
center--of--mass frame the ion beam in a storage ring has similar
physics to that of ions in a Paul trap, we suggest to use {\it absence
of heating} as a diagnostic criterion for the crystalline state of an
ion beam in a storage ring.

%
Before proceeding to the analytical and numerical treatment of absence
of heating, let us introduce an intuitive picture using the classical
phase space of the ion beam. Due to the interplay between the
(effectively time--dependent) confining forces and the Coulomb
interaction between ions the dynamics of the ions is in general
chaotic, resulting in energy diffusion. Nevertheless, the crystalline
state and its vicinity correspond to a region in phase space in which
the dynamics is linear and thus regular. Hence, if linearly stable,
this region can be described as a regular island in phase space
\cite{Ott93,GH83,LL83} in which the trajectories are dynamically
trapped. As a consequence, the time--averaged gain in energy or action
is zero. This is equivalent with the absence of heating. Because of
the possibility of dynamic trapping of phase--space trajectories in
regular islands, it is now clear on physical grounds why ion crystals
can survive even in the absence of a cooling mechanism. We call these
crystals {\it stable}. Thus, we formulate the following diagnostic
criterion for beam crystallization: {\it A beam in a storage ring is
crystallized if no heating occurs following the shut--down of all
cooling devices.} Our criterion works for stable crystals. It is
important to emphasize that the simplest possible ion crystal in a
storage ring, the linear chain
\cite{Hab88,HS90}, is of this type if the density is low (see discussion 
below). It is equally important to emphasize the possibility of {\it
unstable} crystals for which our criterion does not work. The
phase--space trajectory of an unstable crystal is linearly unstable
near the crystalline state in the absence of cooling. A crystalline
state may nevertheless exist if the cooling is strong enough to
overcome the dynamical instability. The resulting crystal is called
unstable since in the absence of cooling the linear instability leads
to heating and melting of the crystal. In this paper we focus our
attention exclusively on the physics of stable crystals deferring
discussion of unstable crystals to a separate publication
\cite{PBinp}.

%
In the remainder of this paper we discuss the physics, the validity
and the limits of applicability of the proposed beam--crystallization
criterion. We illustrate the discussion with the results of numerical
simulations of a model that captures the essence of the physics of
crystallization of very low density beams. The model consists of $N$
ions subject to the time--dependent Hamiltonian
\begin{equation}
  H = K + V_{\rm int} + V_{\rm conf}
  \label{eq:H}
\end{equation}
where
\begin{equation}
  K = \sum_{i=1}^{N} \frac{\vec{P}_i^2}{2m}
  \label{eq:H1}
\end{equation}
is the kinetic energy,
\begin{equation}
  V_{\rm int} = \sum_{1 \leq j < i \leq N}
  \frac{{\cal Z}^2 e^2}{4 \pi \epsilon_0}
  \frac{1}{\left| \vec{R_i} - \vec{R_j} \right|}
  \label{eq:H2}
\end{equation}
is the Coulomb interaction potential, and
\begin{equation}
  V_{\rm conf} = \sum_{i=1}^{N} \left[ 
                 \Phi_0 \cos (\Omega t)(X_i^2 - Y_i^2) -
                 \Psi_0 (X_i^2 + Y_i^2 - 2 Z_i^2) \right]
  \label{eq:H3}
\end{equation}
is the confining potential. Here $\vec R_i=(X_i,Y_i,Z_i)$ are the
Cartesian coordinates of the $i$th ion in the rest frame of the beam,
$\vec P_i$ are its momenta, $m$ denotes the mass of the ions, ${\cal
Z}e$ is their charge, $\Omega$ is the frequency of the confining
fields as seen from the rest frame of the moving ion and $\Phi_0$,
$\Psi_0$ are positive constant parameters. The coordinate system is
oriented such that the $Z$ axis corresponds to the beam axis and the
$X,Y$ plane is orthogonal to the beam. The confining potential $V_{\rm
conf}$ contains a dynamic (time dependent) strongly focusing part in
the transverse $X,Y$ directions that models the action of the
quadrupole focusing magnets, and a bucket--like static part that
models the bunching along the beam direction ($Z$-axis). The
similarity of (\ref{eq:H})--(\ref{eq:H3}) to the Hamiltonian of an $N$
particle Paul trap is apparent. It allows us to draw useful analogies
from the well--developed field of ion traps. The time--dependent part
of the confining potential $V_{\rm conf}$ gives rise to the
micromotion \cite{Deh67} of the beam particles. Away from the
crystalline configuration the micromotion results in fast heating of
the beam, a phenomenon that is completely analogous to the
radio--frequency (rf) heating of ion clouds in a Paul trap
\cite{Deh67}.

%
The equations of motion derived from (\ref{eq:H}) can be conveniently
scaled and are given by:
\begin{eqnarray}
  \ddot{x_i}
  & = &
  -\gamma_x \dot{x_i} + 
  \lambda_0 \sum_{j \neq i, j=1}^{N}
    \frac{(x_i - x_j)}{\left| \vec{r_i} - \vec{r_j} \right|^3} +
  \left[ a - 2q \cos (2 \tau) \right] x_i \; ,
  \label{eq:eqomx} \\
  \ddot{y_i}
  & = &
  -\gamma_y \dot{y_i} + 
  \lambda_0 \sum_{j \neq i, j=1}^{N} 
    \frac{(y_i - y_j)}{\left| \vec{r_i} - \vec{r_j} \right|^3} +
  \left[ a + 2q \cos (2 \tau) \right] y_i \; ,
  \label{eq:eqomy} \\
  \ddot{z_i}
  & = &
  -\gamma_z \dot{z_i} + 
  \lambda_0 \sum_{j \neq i, j=1}^{N} 
    \frac{(z_i - z_j)}{\left| \vec{r_i} - \vec{r_j} \right|^3} -
  2 a z_i \; ,
  \label{eq:eqomz}
\end{eqnarray}
where we used the following definitions:
\begin{equation}
  \Omega t \equiv 2 \tau \; , \; \;
  \vec{R}_i \equiv l_0 \vec{r}_i \; , \; \;
  \lambda_0 \equiv \frac{{\cal Z}^2 e^2}
                         {\pi \epsilon_0 m \Omega^2 l_0^3} \; , \; \;
  a \equiv  \frac{8 \Psi_0}{m \Omega^2} \; , \; \;
  2q \equiv \frac{8 \Phi_0}{m \Omega^2} .
\end{equation}
In the above equations we added a (possibly anisotropic) damping
effected by the damping constants $\gamma_x$, $\gamma_y$, $\gamma_z$
to model the laser cooling.

%
In our computations we model $N=5$ ions of $^{24}$Mg$^+$ with $q=0.2$
and $a=q^2/64$. The external frequency $\Omega$ is $2\pi\times 3$ MHz
and the length scale $l_0$ is chosen to be 1 $\mu$m. The damping
serves as our control parameter. With this choice of parameters the
typical spatial extension of the ion ensemble in $z$ direction is
about four times the extension in the $x,y$ directions. In other
words, our parameters are chosen such that the ions experience a
focusing force that is about four times stronger in the $x,y$
directions compared to the $z$ direction. For this choice of
parameters the linear chain is the lowest energy crystalline
configuration of our model system
\cite{Hab88,HS90}. The integration of the equations of motion
(\ref{eq:eqomx})--(\ref{eq:eqomz}) was performed using a
variable--order, variable--step Adams method \cite{NAG90}.

%
Although our model of five ions seems like a caricature of a real beam
in a storage ring, it captures the essence of cooling and
crystallization of low-density beams in strongly focusing machines.
An indication to this effect is the following observation. When we
simulated anisotropic cooling, i.e., $\gamma_x=\gamma_y=0$, $\gamma_z
= \gamma>0$, we faced the same problem as the experiments: with
anisotropic cooling it is very difficult to create Coulomb crystals
with reasonable choices of $\gamma$. Moreover, we think of our model
as describing a typical section of the beam rather than a system
consisting of a small number of ions. This is substantiated by the
fact that near the linear crystal state essentially only the nearest
neighbor Coulomb interactions are important, since the Coulomb force
reduces like the square of the distance. Hence the number of total
ions is not of primary importance. We note that the static confining
potential is suitable for bunched beams, while periodic boundary
conditions that preserve linear density are suitable for coasting
beams. We emphasize that our calculations are microscopic and do not
suffer from any further approximations beyond the choice of the model
Hamiltonian (\ref{eq:H}).

%
In order to keep things as simple as possible, and since the
applicability of our crystallization criterion does not depend on the
type of cooling scheme used to produce a crystallized beam, we
illustrate the criterion and the physics of beam crystallization with
the help of isotropic cooling, i.e.\ $\gamma_x = \gamma_y = \gamma_z =
\gamma > 0$.

%
Many of our numerical results reported below are discussed within the
framework of ion temperature $T$. The temperature here is only a
convenient way of expressing the experimentally measured mean--square
of the ion velocity. We follow the convention used by the
experimentalists and use the standard thermodynamic relation $\langle
( \dot{\vec{R}} - \langle \dot{\vec{R}} \rangle )^2 \rangle \equiv 3
k_{\rm B} T / m$, where $\langle \cdots \rangle$ denote averaging over
ions. We thus avoid all questions of the thermodynamic relevance of
defining a temperature for five particles as well as the question of
how to define a temperature in the presence of a strong coherent
drive.

\begin{figure}[tp]
  \centerline{\psfig{figure=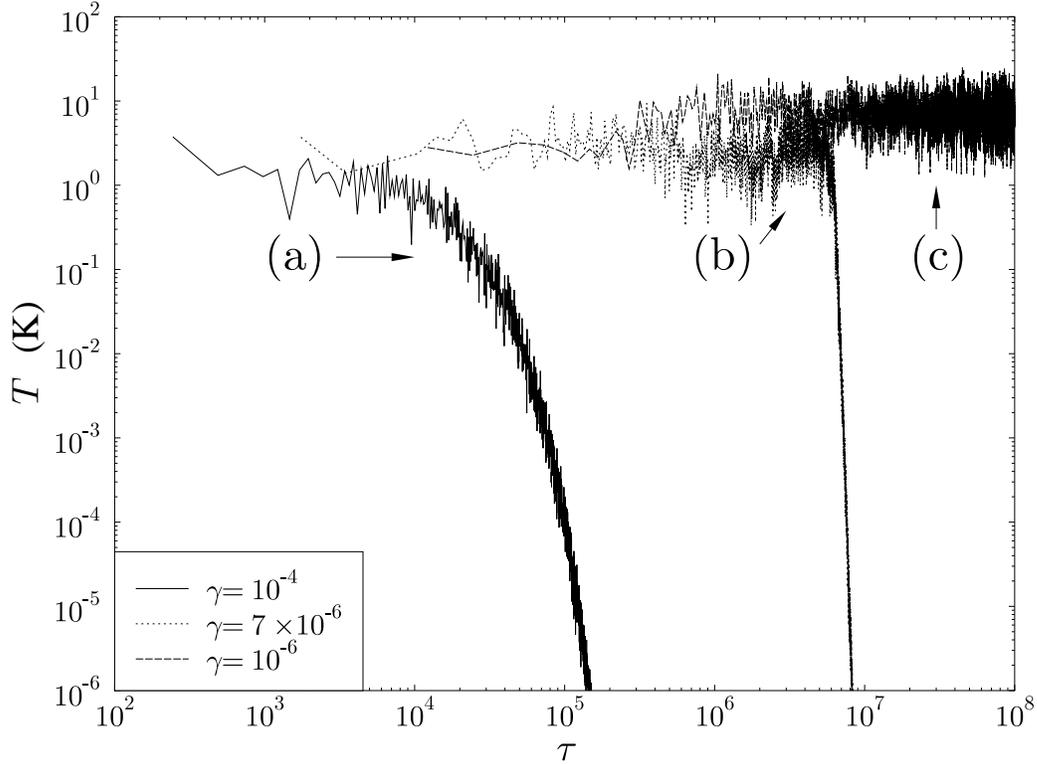,width=15cm}}

  \caption{Three different scenarios for the time evolution of an
           initially hot state (cloud) of ions: (a) Immediate
           crystallization ($\gamma=10^{-4}$, strong damping regime),
           (b) intermittency ($\gamma=7\times 10^{-6}$, intermediate
           damping regime) and (c) persistent cloud ($\gamma=10^{-6}$,
           weak damping regime).}

  \label{fig:temp-3-cases}
\end{figure}

\begin{figure}[tp]
  \centerline{\psfig{figure=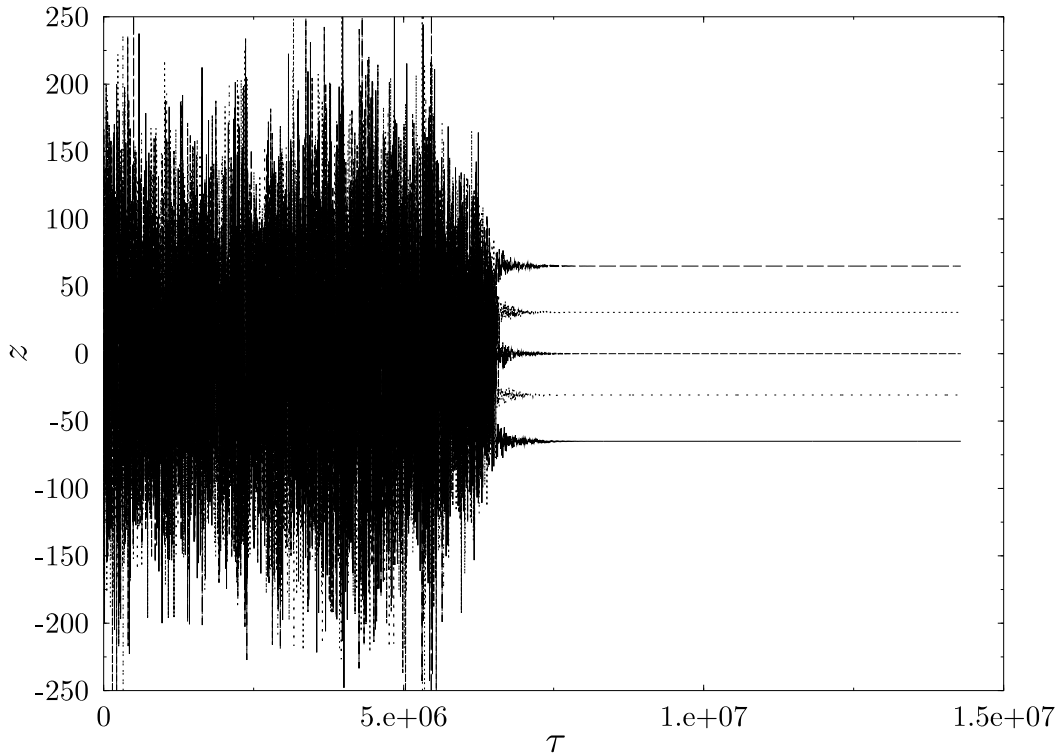,width=15cm}}

  \caption{Time dependence of the $z$ coordinates of the five ions of
           the intermittent case (b) of figure
           \protect\ref{fig:temp-3-cases}. The ions crystallize out of
           a transiently chaotic state at $\tau \approx 6.5 \times
           10^6$ and arrange themselves regularly spaced in a
           crystallized linear chain. The lattice spacing in the
           crystalline state is $\Delta_z \approx 30 \mu$m.}

  \label{fig:ions-z-inter}
\end{figure}

%
Before embarking on the central point of diagnosing the crystal state,
we present some relevant results relating to the cooling and
crystallization process itself. We show that, depending on the cooling
strength $\gamma$, an initially prepared generic hot ion cloud
\cite{BCPQS88,BKQW89} evolves according to three qualitatively different
scenarios. In figure \ref{fig:temp-3-cases} we show the temperatures
of our five ion ensemble as a function of (scaled) time $\tau$, and
for three different values of $\gamma$. Curve (a) shows the
temperature for $\gamma=10^{-4}$. The temperature decays exponentially
($\sim\exp(-\gamma\tau)$) right from the beginning of the
simulation. This is the strong damping regime. Curve (b) of figure
\ref{fig:temp-3-cases} shows a qualitatively different behavior for an
intermediate choice of the cooling strength $\gamma=7\times 10^{-6}$.
The beam is in the state of a hot cloud over a substantial period of
time (much longer than $1/\gamma$). It collapses suddenly (at an
exponential rate) to zero temperature from some time on. This behavior
is called intermittent \cite{HB94}. Curve (c) of figure
\ref{fig:temp-3-cases}, for $\gamma=10^{-6}$, shows a hot beam which
remains hot over the total time interval of our simulations
($\tau_{\rm max} = 10^8 = 100/\gamma$) although the cooling is
constantly switched on. In this case the cooling is balanced by the rf
heating. The heating is due to Coulomb collisions (intra-beam
scattering) in the presence of the time--dependent fields in
(\ref{eq:H}). This mechanism is also active in ion traps
\cite{BHDRH90} and has been identified as the dominant heating
mechanism in storage ring beams \cite{HG95}. In both the weak and the
intermediate cooling regimes the end result of the simulations, i.e.,
the zero temperature state, is the linear chain. This is illustrated
in figure
\ref{fig:ions-z-inter}. It shows the $z$-coordinates of our five-ion
ensemble corresponding to curve (b) of figure
\ref{fig:temp-3-cases}. After a chaotic transient lasting for about $0
\leq \tau \leq 6 \times 10^6$ we see a sudden crystallization of the
five ions into stationary positions approximately equi-spaced on the
$z$ axis. Inspection of the $x$ and $y$ coordinates shows $x_i \approx
0$, $y_i \approx 0$. Thus, the crystalline state corresponds to a
linear chain. In our simulations, as seen from figure
\ref{fig:ions-z-inter}, the spacing of the ions is $\Delta_z
\approx 30 \mu$m. The linear chain is the natural crystalline
configuration for very low density crystalline beams. It does not
suffer from the problem of shear heating \cite{RS86}, a serious
obstacle on the way to three-dimensionally crystallized beams in
presently existing accelerator storage rings. Thus, establishing the
linear chain is an important corner stone on the way to more
complicated crystal configurations. Chaotic transients similar to the
one shown in figure \ref{fig:temp-3-cases} were described earlier by
Hoffnagle and Brewer \cite{HB94} in connection with the two-ion Paul
trap.

\begin{figure}[tp]
  \centerline{\psfig{figure=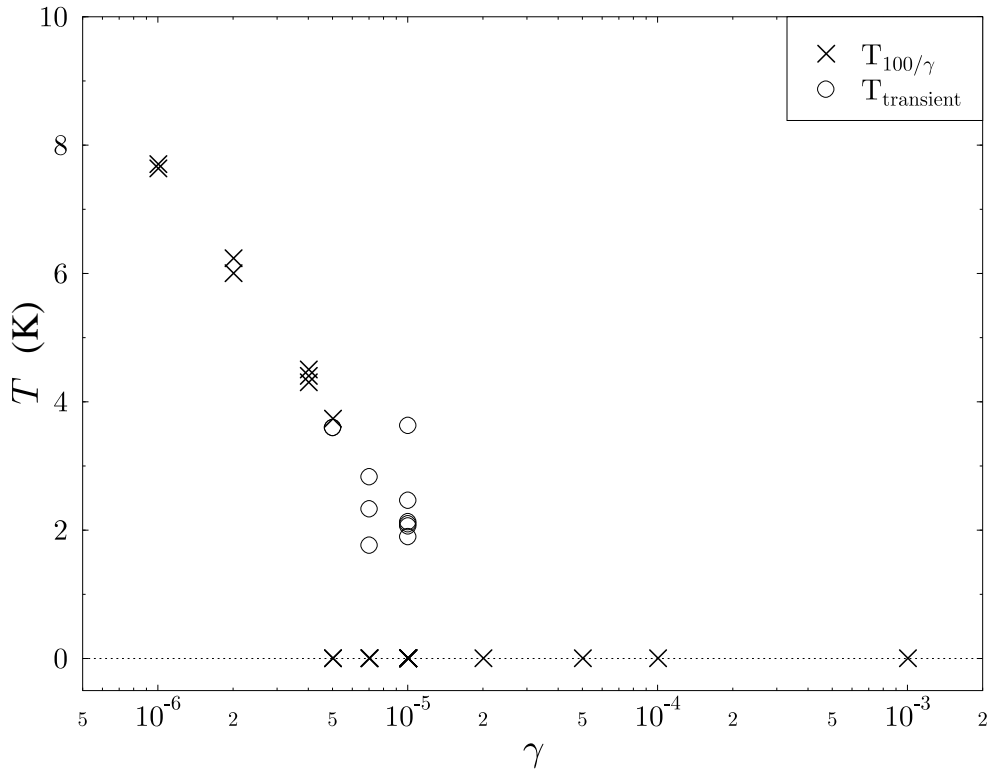,width=15cm}}

  \caption{Beam temperatures as a function of cooling strength
           $\gamma$. Crosses: Final beam temperatures after a run time
           of $\tau = 100/\gamma$. Circles: Beam temperature shortly
           before the onset of the sharp crystallization transition in
           the intermittent regime.}

  \label{fig:temps-gamma}
\end{figure}

%
The statistical significance of the results shown in figures
\ref{fig:temp-3-cases} and \ref{fig:ions-z-inter} is substantiated 
in figure \ref{fig:temps-gamma}. It gives a global view of the cooling
process by showing the final beam temperatures after $\tau_{\rm max} =
100 / \gamma$ for a wide range of $\gamma$ values and a few initial
conditions each. It shows also the temperatures in the cloud state in
the intermittent cases. We observe that for small cooling power the
beam remains in a hot state over the entire observation time $0 \leq
\tau \leq \tau_{\rm max}$. The temperature of the hot state decreases
with increasing cooling strength. For strong cooling, crystallization
takes place immediately. For intermediate cooling power there is
intermittency, but the overall transition from a cloud to a crystal
state is sharp as a function of $\gamma$.

\begin{figure}[tp]
  \centerline{\psfig{figure=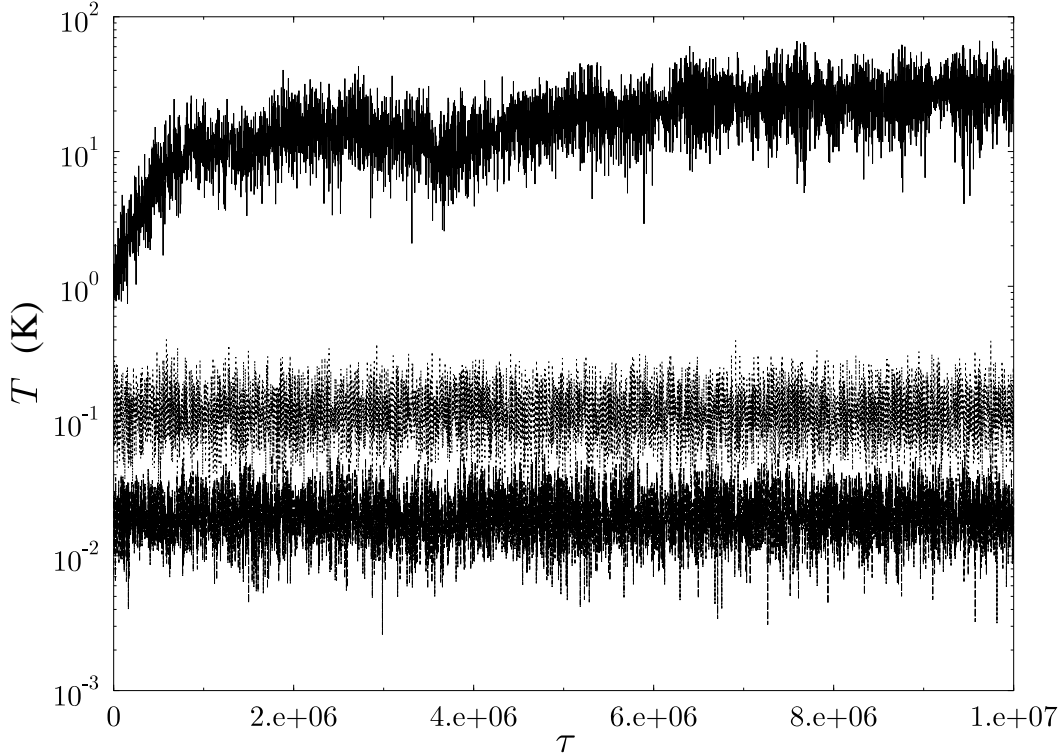,width=15cm}}

  \caption{Time dependence of beam temperatures for three different
           initial conditions close to the crystalline configuration
           with no cooling $(\gamma = 0)$. Upper curve: Initial
           temperature is $6.7 \times 10^{-1}$ K, middle curve:
           Initial temperature is $6.7 \times 10^{-2}$ K, lower curve:
           Initial temperature is $6.7 \times 10^{-3}$ K. The upper
           curve shows strong heating. It does not correspond to a
           crystalline state. The two cases with lower initial
           temperatures show absence of heating. They correspond to a
           crystallized beam according to the proposed diagnostic
           criterion.}

  \label{fig:temp-hyst}

\end{figure}

%
We now turn to the key point of this paper: The absence of heating of
ion crystals. Absence of heating serves as a simple diagnostic tool
for the crystalline state of ion beams. We shall focus here
exclusively on linear crystals. In order to demonstrate the absence of
heating, we picked three different initial conditions of the ion beam
close to the linear crystalline state and integrated the equations of
motion with the damping switched off ($\gamma=0$). In figure
\ref{fig:temp-hyst} we plot the resulting temperatures as a function
of the time for three representative cases, in which the initial
temperatures were $6.7 \times 10^{-3}$ K, $6.7 \times 10^{-2}$ K and
$6.7 \times 10^{-1}$ K. The integration was carried out over the time
interval $0 \leq \tau \leq 10^7$ which corresponds to $\approx 3
\times 10^6$ cycles of the focusing force. For the two cases at lower
temperature, we clearly observe the absence of heating. This means
that even when there is no cooling, the beam remains very close to the
linear crystalline state for a very long time. As expected, when the
initial temperature is large enough (about 1 K in our case) the
crystal melts quickly due to rf heating and the temperature increases
dramatically. After an initial fast ``blow--up'' phase the heating
continues on a slower scale. The initial blow--up shown in figure
\ref{fig:temp-hyst} strongly resembles the beam blow-up shown in
\cite{MGGHS96} after switch--off of the electron cooler. From figure
\ref{fig:temp-hyst} we conclude that if the linear crystal is
cold enough, then it can maintain itself even if the cooling power is
switched off.

%
A few comments are in order now. First, as discussed above, our
criterion fails to identify unstable crystals. Second, especially if
the cooling process is rapid, we may produce a disordered, ``glassy''
state without geometric order that may be stable in the absence of
cooling. Thus our criterion would wrongly identify a glass as a
crystal. In general, therefore, applying the strictest standards, our
criterion is neither necessary nor sufficient for the proper
identification of geometrically ordered crystalline states. However if
we restrict ourselves to low--density ion beams, which is the current
experimental trend, then we are in much better shape. Indeed, if the
ion density is small enough, the beam will crystallize to a linear
chain, which excludes glassy states. This is substantiated by the work
of Hasse and Schiffer \cite{HS90} and by the work of Habs \cite{Hab88}
as well as by our numerical simulations. Quantitatively it requires
$\lambda \equiv \lambda_0 / \Delta_z^3 \lesssim 1$. Moreover, a
detailed analytical investigation which we carried out and will appear
elsewhere \cite{PBinp} indicates that the linear crystal is stable in
the absence of cooling, provided that in the two-dimensional parameter
space of the Mathieu equation the interval $([a,a+4.21\lambda],q)$ is
contained in a stable region. This is the case if $\Delta_z$ is large
enough, i.e., for very low density beams. Hence the above analysis
indicates that for small values of the parameter $\lambda$ our
no-heating criterion is both necessary and sufficient for the
identification of a linear crystal. Let us note that in our
simulations $\lambda\approx 2\times 10^{-3}$. Indeed, we never
encountered a glassy state and our linear crystal was stable even
without cooling in full agreement with the above discussion.

%
In practice our diagnostic criterion may be applied according to the
following three steps:
\begin{enumerate}

  \item We start with a hot beam.

  \item The cooling devices are switched on (e.g.\ electron and/or
        laser cooling) resulting in a cold beam whose state (crystal
        or not) is to be determined.

  \item All cooling devices are switched off.
\end{enumerate}
If the beam is indeed crystallized, no heating will be observed. The
beam remains crystallized. If the beam is just cold, but not
crystallized, heating is observed after shut--down of the cooling
devices. We emphasize that the shut-down should occur slowly, not
abruptly. This is because in our simulations we noticed that a sudden
switch--off of the damping results in a phase--jump in the time
dependence of the ion trajectories leading to instantaneous heating of
the system that may be strong enough to disrupt the crystalline
state. However, there is no problem if the shut--down is slow on the
scale of the micromotion of the beam particles.

%
A last comment concerns the beam observation time scale after
shut--down of the cooling devices. It is clear that even if the beam
was crystallized, it will heat slowly due to, e.g.\ fluctuations in
the confining fields and the ambient thermal radiation. When we talk
about heating after shut--down of the cooling devices we mean the fast
rf heating due to the time--dependent confining fields in
(\ref{eq:H}). Since the noise processes are much slower than rf
heating, there should be no problem to separate the two mechanisms
experimentally.

%
To summarize, in this paper we proposed a beam diagnostic criterion
for deciding whether an ion beam in a storage ring is crystallized or
not. The criterion is simple to apply and does not require any
technical installations that are not already present in existing
storage rings. With the help of model calculations we demonstrated
that the criterion works well for low--density crystallized ion
chains. The criterion has a limited range of validity, but since we
expect that the first crystalline geometry achieved in a storage ring
will be the linear chain, the crystallization criterion may play an
important role in proving experimentally the presence of a
crystallized beam in a storage ring.

%
The authors acknowledge fruitful discussions with R.\ Grimm and with
M.\ Weidem\"uller. H.\ P.\ is grateful for a MINERVA postdoctoral
scholarship. R.\ B.\ is grateful for financial support by the Deutsche
Forschungsgemeinschaft (SFB 276).




\end{document}